\documentclass{scrartcl}
\usepackage[affil-it,auth-lg]{authblk}

\usepackage[latin1]{inputenc}
\usepackage{amsmath}
\usepackage{amsfonts}
\usepackage{amssymb}
\usepackage{bm,amsmath,amssymb,amsfonts,latexsym,psfrag}
\usepackage{bm}
\usepackage{amssymb}
 \usepackage{array}
\usepackage{mathtools}
\usepackage{amsthm}
\usepackage{newlfont}
\usepackage{esdiff}
\usepackage{multirow}

\usepackage{graphicx}
\usepackage{parskip}
\usepackage{graphicx}
\usepackage{url}                
\usepackage{bm,amsmath,amssymb,amsfonts,latexsym,psfrag}
\usepackage[small,it]{caption}
 \usepackage{amssymb}
 \usepackage{enumerate}
 \usepackage{color}
 \let\origthanks\thanks\renewcommand\thanks[1]{\begingroup\let\rlap\relax\origthanks{#1}\endgroup}

\newcommand{\beq}{\begin{equation}}
\newcommand{\eeq}{\end{equation}}
\newcommand{\bea}{\begin{eqnarray}}
\newcommand{\eea}{\end{eqnarray}}

\newcommand{\LP}{L_{\mathrm{P}}}
\newcommand{\MP}{M_{\mathrm{P}}}
\newcommand{\T}{T_{\mathrm{H}}}
\newcommand{\rM}{r_{\mathrm{M}}}

\title{Holographic screens in ultraviolet self-complete quantum gravity}

\author[a,b]{Piero Nicolini\footnote{\textit{E-mail:} \texttt{nicolini@fias.uni-frankfurt.de}}}
\author[c]{Euro Spallucci\footnote{\textit{E-mail:} \texttt{spallucci@ts.infn.it}}}
\affil[a]{Frankfurt Institute for Advanced Studies (FIAS), Ruth-Moufang-Strasse 1, D-60438 Frankfurt am Main, Germany}
\affil[b]{Institut f\"{u}r Theoretische Physik, Johann Wolfgang Goethe-Universit\"{a}t,  Max-von-Laue-Strasse 1, D-60438 Frankfurt am Main, Germany}
\affil[c]{Dipartimento di Fisica, Sezione di Fisica Teorica, Universit\`a degli Studi di Trieste and INFN, Sezione di Trieste, Strada Costiera 11, I-34151 Trieste, Italy}

\date{\large\today}

\begin{document}

\maketitle

\begin{abstract}
In this paper we study the geometry and the thermodynamics of a \emph{holographic screen}
in the framework of the ultraviolet self-complete quantum gravity. 
To achieve this goal we construct a new static, neutral, non-rotating black hole metric, whose outer (event) 
horizon coincides with the surface of the screen. The space-time admits an extremal configuration corresponding 
to the minimal holographic screen and having both mass and radius equalling the Planck units.
We identify this object as the space-time fundamental building block, whose interior is physically 
unaccessible and cannot be probed even during the Hawking evaporation terminal phase.  
In agreement with the holographic principle, relevant processes take place on the screen surface.
The area quantization leads to a discrete mass spectrum. An analysis of the entropy shows that 
the minimal holographic screen can store only one byte of information while in the thermodynamic 
limit the area law is corrected by a logarithmic term. 

\end{abstract}


\section{Introduction}
\begin{figure}[t!]
\begin{center}
\includegraphics[width=0.6\textwidth]{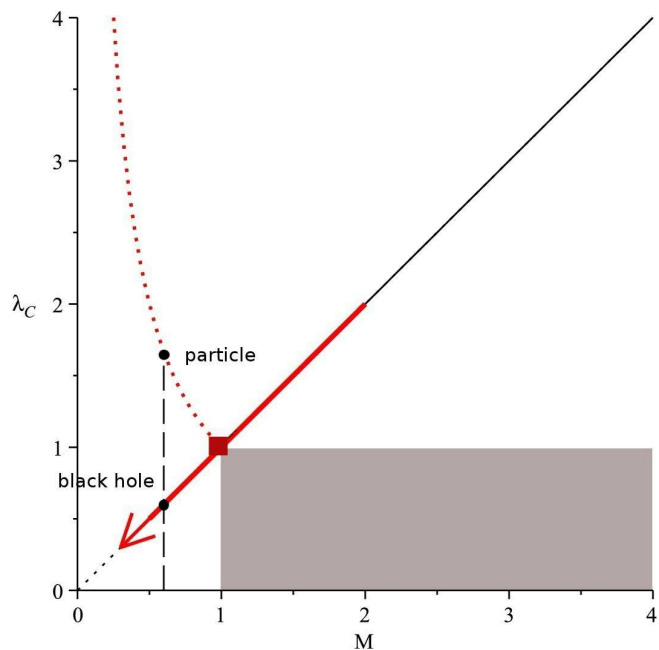}
\caption{\label{fig:bhdecay}The dotted and the solid curves represent the particle Compton wave length $\lambda_C$ and the Schwarzschild radius as a function of the energy $M$ in Planck units (quantities are  rescaled). 
The squared bullet is the Planck scale. The grey area of the diagram is actually excluded, meaning that a particle cannot be compressed at distances smaller than the Planck length: at trans-Planckian energy only black hole form.  
The arrow shows the inadequacy of the Schwarzschild metric: black holes have no lower mass bounds, can have size smaller than the Planck length  and can expose the curvature singularity by decaying through the Hawking process. 
}
\end{center}
\end{figure}

``Quantum gravity'' is the common tag for any attempt to reconcile gravity and quantum mechanics.
Since the early proposals by Wheeler \cite{jaw,jaw2} and deWitt \cite{bdw}, up to the recent ultraviolet (UV) self-complete scenario \cite{gia},
the diverse formulations of a would be quantum theory of gravity have shown a common feature, \textit{i.e.}
a fundamental length/energy scale where the smooth manifold model of spacetime breaks down.
Let us refer to this scale as the ``Planck scale'' irrespectively whether it is $10^{19}\, \mathrm{GeV} $
or, $10-10^2\, \mathrm{TeV} $. The very concept of distance becomes physically meaningless at the Planck
scale and spacetime ``evaporates'' into something different, a sort of ``foamy'' structure, a spin network, 
a fractal dust, \textit{etc.}, according with the chosen model \cite{garay}.
As a matter of fact, one of the most powerful frameworks for describing
the Planckian phase of gravity is definitely (Super)String Theory. The price to pay to have
a perturbatively finite, anomaly-free quantum theory is to give up the very idea  of
point-like building blocks of matter, and replace them with one-dimensional vibrating strings.
As there does not exist any physical object smaller than a string, there is no physical ways
to probe distances smaller than the length of the string itself. 
In this regard two properties of fundamental strings are worth mentioning:
\begin{itemize}
 \item string excitations correspond to different mass and spin ``particle'' states;
  \item highly excited strings share various physical properties with black holes.
\end{itemize}
Thus, we infer that string theory provides a bridge between particle-like objects and black holes 
(see for instance \cite{hp97}).  
However, it is important to remark that while the  Compton wavelength of a particle-type excitation
decreases by increasing the mass, the Schwarzschild radius of a black hole increases with its mass.
Thus, the first tenet of high energy particle physics, which is ``higher the energy shorter the distance'',
breaks down when gravity comes into play and turns a ``particle'' into a black hole. 
The above remark is  the foundation of the UV self-complete quantum gravity scenario, where
the Planckian and sub-Planckian length scales are permanently shielded from observation due to
 the production of black hole excitations at Planck energy scattering \cite{gia2}. Accordingly the Planck scale assumes the additional meaning of scale at which matter undergoes 
a transition between its two admissible ``phases'', \textit{i.e.}, the particle phase and the black hole phase \cite{gia3,gia4,gia5}. 
From this perspective,  trans-Planckian physics is dominated by larger and larger black hole configurations.  
It follows that  only black holes  larger, or at most equal to Planck size objects, can self-consistently fit into this scheme.  
However, classical black hole solutions do not fulfill this requirement, \textit{i.e.} the existence of a lower bound for their mass and size (see Fig. \ref{fig:bhdecay}). 

A first attempt to overcome this limitation is offered by the noncommutative geometry inspired solutions of the Einstein equations \cite{Nicolini:2008aj}. The latter are a family of regular black holes which span all possible combinations of parameters, such mass \cite{trieste1}, charge \cite{trieste3}, angular momentum \cite{trieste6,trieste7}.   In addition such regular geometries admits a variety of complementary gravitational configurations such as traversable wormholes \cite{remo}, dirty black holes \cite{trieste5}, dilaton gravity black holes \cite{dgbh} and collapsing matter shells \cite{shell}. Recently this family of black holes has been recognized as viable solutions of non-local gravity \cite{mmn,piero2012}, \textit{i.e.}, a set of theories exhibiting an infinite number of derivative terms of the curvature scalar \cite{nl1,nl2,nl3} in place of the mere Ricci scalar as in the standard  Einstein-Hilbert action. More importantly extensions of noncommutative geometry inspired metrics to the higher dimensional scenario \cite{pheno1,trieste4} are currently under scrutiny at the LHC for their unconventional phenomenology \cite{pheno3}: specifically the terascale black holes described by such regular metrics tend to have a slower evaporation rate \cite{pheno2} and emit only soft particles  mainly on the four dimensional brane \cite{pheno4}. 
A characteristic feature of this  type of  solutions is that the minimum size 
configuration is given by the extremal black hole configuration which exists even in the neutral non-spinning case \cite{trieste2,colombia,india1}.  This fact automatically implies a minimum energy for black hole production in particle collisions \cite{jpe} without any further need of correcting formulas of cross sections with \textit{ad hoc}  threshold functions. Extremal configurations play a crucial role in the physics of the decaying deSitter universe via the nucleation of microscopic black holes. It has been shown that Planck size noncommutative inspired black holes might have been copiously produced during inflationary epochs  \cite{nucleation}. This fact has further phenomenological repercussions: being stable, non-interacting objects, extremal black holes turn out to be a reliable candidate for  dark matter component. On the theoretical side, extremal configurations in the presence of a negative cosmological term can provide a short scale completion of the Hawking-Page diagram which switches to a more realistic Van der Waals phase diagram \cite{vdw}.

Extremal configurations can be either descend from the introduction of a fundamental length in the line element and  can alternatively be interpreted as a phenomenological input from quantum gravity: in the latter case  it has been shown that such extremal black holes fit pretty well in the UV self-complete scenario providing a stable, minimum
size, probe at the transition point between particles and black holes \cite{sa11}.
 
In this paper we want to take a step further in the realization of this program by avoiding the introduction 
of an additional principle to justify the presence 
of a minimal length, rather we demand  the radius of a Planck size extremal black hole to provide the natural UV cut-off 
of a quantum spacetime.
In this framework gravity is expected to be self-regular in the sense that the actual regulator
 cutting off sub-Planckian length scales is given in terms of the gravitational coupling constant,
\textit{i.e.}  $\sqrt{G}=\LP$.    
The paper is organized as follows. In Section II we derive a black hole metric,  consistent with the above discussion and the concept of holographic screen. 
The latter coincides with the outer horizon of the black hole whose mass spectrum is bounded from below by the mass of the extremal configuration equalling the Planck mass. 
Once trans-Planckian length scales
are cut-off, the ``interior" of the black hole loses its physical meaning in the sense that all the relevant degrees 
of freedom are necessarily located on the horizon itself. 
In Section III we discuss the thermodynamics of the screen. We find that  the area law is modified
by logarithmic corrections and that there exists a minimal holographic screen with zero thermodynamic entropy.
Finally we propose an ``holographic quantization'' scheme where the area of the extremal configuration provides the 
quantum of surface. 
In Section IV we offer to the reader a brief summary of the main results of this work.

\section{Self-regular holographic screen}
A simple but intriguing model of singularity-free black hole
has been ``guessed'' in \cite{Hayward:2005gi}, in the sense that the metric was assigned as  an in-put for the Einstein equations.
Sometimes this inverted procedure is called ``engineering'' because the actual source term of  field equations is not known 
a priori.
The distinctive feature of the solution is the presence in the line element
of a free parameter with dimension of a length, acting as a short distance regulator for the
spacetime curvature, allowing a  safe investigation of back-reaction effects of the Hawking radiation.
In  \cite{Spallucci:2012xi} an higher dimensional extension of this model has been proposed; it was also
shown that, by a numerical rescaling of the short-distance regulator,  it is possible to identify 
this fundamental length scale with the radius of the extremal configuration.
With hindsight, we are going to  take a step forward to improve this inverse procedure.
Specifically, we want to follow the ``\emph{direct way}'' by building up a consistent source for Einstein equations:  
we introduce a physically motivated energy momentum 
tensor which allows for transitions between particle-like objects and black holes as consistently required by UV 
self-complete quantum gravity.

We start from the energy density for a point-particle in spherical coordinates as
\begin{equation}
\rho_p\left(\, r\,\right)=\frac{M}{4\pi r^2}\, \delta\left(\, r\,\right),
\label{rosing}
\end{equation}
where $\delta\left(\, r\,\right)$ is the Dirac delta.
The energy distribution (\ref{rosing}) implies a black hole for any value of mass $M$ even for sub-planckian
values where one expects just particles. 
Before proceeding, we would like to recall  that a Dirac delta function can be represented as the derivative of a 
Heaviside step-function $\Theta$
\begin{equation}
 \delta\left(\, r\,\right)=\frac{d}{dr}\Theta\left(\, r\,\right).
\end{equation}
Against this background, we want to accommodate both particles and black holes by a suitable modification of the  energy 
distribution in order to overcome the 
ambiguities of the Schwarzschild metric in the sub/trans-Planckian regimes (see also Fig. \ref{fig:bhdecay}).
This can be done by considering a ``~smooth~'' function $h(r)$ in place of the Heaviside step
\begin{equation}
\Theta\left(\, r\,\right)\longrightarrow h\left(\,r\,\right).
\end{equation}
The new profile $\rho(r)$ of the energy density is defined through $h(r)$ by the relation 
\begin{eqnarray}
\rho(r)=   \frac{M}{4\pi r^2}\,  \frac{d}{dr}  \ h(r) \equiv T_0^0.
\end{eqnarray}
By means of the conservation equation $\nabla_{\mu}T^{\mu\nu}=0$ one can determine the remaining
components of the stress tensor, which turns to be out of the form 
\begin{equation}
T_{\mu}^{\nu}=
\mathrm{diag\left(-\rho,\,\mathit{p}_{\mathit{r}},\,\mathit{p}_{\perp,}\,\mathit{p}_{\perp}\right)}
\label{timunu}
\end{equation}
The condition for the metric coefficients $g_{00}=-g_{11}^{-1}$ determines the equation of state,
namely the relation between the energy density and the radial pressure,
$\mathit{p_{\mathrm{\mathit{r}}}=-\rho}$. The angular
pressure is specified by the conservation of the stress tensor and
reads $p_{\perp}=p_{r}+\frac{r}{2}\partial_{r}p_{r}$.

By plugging the tensor (\ref{timunu}) in Einstein equations, one finds that the metric reads $(G=1)$
\begin{eqnarray}
ds^{2}= &&-\left( 1-\frac{2m\left(r\right)}{r}\right)\, dt^{2}\nonumber\\
        &&+\left( 1-\frac{2m\left( r\right)}{r}\right)^{-1}\, dr^{2}+r^{2}d\Omega^{2}\ ,
\end{eqnarray}
with
\begin{equation}
m(r)= 4\pi \int^r dr'(r')^{2}\, \rho\left(\ r^\prime\,\right)\ 
\end{equation}
At large distances $r\gg\LP$, the above energy density has to quickly vanish, \textit{i.e.} $\rho(r)\to 0$ in order 
to match the ``~vacuum~'' Schwarzschild metric.
Conversely, at shorter scales $r\gtrsim \LP$, the density $\rho(r)$ (and accordingly $h(r)$) has to depart from the 
point-particle profile in order to fulfil the following  
requirements: 
\begin{enumerate}[i)]
\item \label{uno} no curvature singularity in the origin; 
\item \label{tre} \textit{self-implementation} of a characteristic scale $l_{0}$ in the spacetime geometry 
by means of the radius of the extremal 
configuration $r_{0}$, \textit{i.e.}, $r_{0}=l_{0}$. 
\end{enumerate}
%
\begin{figure}[t!]
\begin{center}
\includegraphics[width=0.6\textwidth]{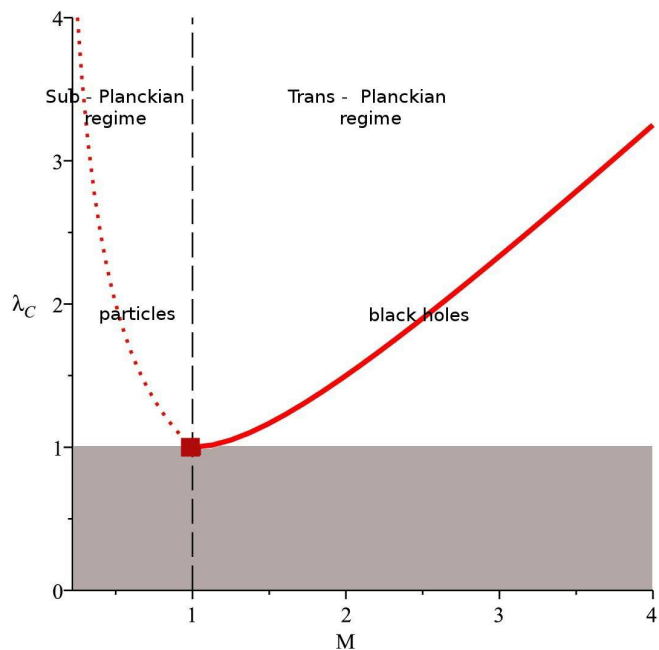}
\caption{\label{fig:hsc}The plot shows a length/energy relation consistent with the self-complete quantum gravity 
arguments in Planck units. Particles (dotted line) and black holes (solid line) cannot probe length shorter than the Planck length. 
The grey area is permanently inaccessible and accordingly represents the minimal spacetime time region  or fundamental 
constituent \textit{i.e.} the ``atom'' the spacetime is supposed to be made of. 
} 
\end{center}
\end{figure}

The latter condition is crucial. For instance noncommutative geometry
inspired black holes \cite{Nicolini:2008aj} are derived by the direct way, they enjoy i), 
but fail to fulfill the condition ii). This
means that the characteristic length scale of the system $l_{0}$
and the extremal configuration radius $r_{0}$ are independent quantities.
Indeed noncommutative geometry is the underlying theory which provides
the scale $l_{0}$ in terms of an ``external'' parameter, namely
the noncommutative parameter $\theta$. In other words one needs to
invoke a principle, like a modification of commutators in quantum
mechanics, or the emergence of a quantum gravity induced fundamental
length to achieve the regularity of the geometry at short scales.
Against this background, we want just to use $r_{0}$ as fundamental
scale, getting rid of any $l_{0}$ as emerging from any theory or
principle not included in Einstein field equations. This is a step
forward since it opens the possibility for Einstein gravity to be
self-protected in the ultraviolet regime. To emphasize this point, we introduced the word ``\textit{self}-implementation'' in ii). 
Since there exists actually only 
one additional scale beyond $r_{0}$, \textit{i.e.} the Planck length $\LP=\sqrt{G}$,
or the Planck mass $\MP=1/\sqrt{G}$, we can  implement 
the condition ii) in the most natural way by setting $r_{0}=\LP$ and accordingly $M_0=\MP$, where $M_0\equiv M(r_0)$ 
is the extremal black hole mass. 

Despite the virtues of the above line of reasoning, we feel that the set of conditions i) 
and ii) can be relaxed and a further simplification is possible.   
Having in mind that for extremal black hole configurations the Hawking emission stops we just need to 
find a metric for which only the condition ii) holds. 
This would be enough for completing the program of the UV self-complete quantum gravity by protecting the short distance 
behavior of gravity during the final stages of 
the evaporation process.  In this regard, the resulting extremal black hole  is just the smallest object one can use to  
probe short-distance physics, 
In other words, in the framework of UV self-complete quantum gravity,  \emph{it is not physically meaningful 
to ask about curvature singularity inside the horizon as the very concept of spacetime is no longer defined below this length scale}.
 
According with such a line of reasoning, we can determine the function $h(r)$ by dropping the condition i) and keeping just 
the condition ii). Inside the class of all 
admissible profiles for $h(r)$, the most natural and algebraically compact choice is given by 
\begin{equation}
h\left(\,r\,\right)= 1-\frac{\LP^2}{r^2 + \LP^2}
\end{equation}
A similar procedure has been already used in \cite{jpe} and accounts for the fact that in the presence
of $\LP$ the step cannot be any longer sharp.
Thus, the smeared energy density $\rho(r)$ turns out to be 
\begin{eqnarray}
\rho(r)=  \frac{M}{2\pi r} \, \frac{\LP^2}{\left(\, r^2 +\LP^2\,\right)^2}\label{sorgente}
\end{eqnarray}
As a result we find the following metric which is derived from a stress tensor modeling a particle-black hole
 system (\ref{timunu})
\begin{eqnarray}
ds^{2}=-\left(\,1-\frac{2M\LP^{2}\, r}{ r^2
+\LP^2} \,\right)dt^{2} +\left( \, 1-\frac{2M\LP^{2}\, r}{ r^2+\LP^2  } \, \right)^{-1}dr^{2}+r^{2}d\Omega^{2},
\label{eq:metric-2}
\end{eqnarray}
where the arbitrary constant $M$ is defined  as follows:
\begin{equation}
 M\equiv \frac{1}{2\LP^2\, r_\mathrm{h}}\left(\, r_\mathrm{h}^2 + \LP^2\,\right)\ .\label{mholo}
\end{equation}
We give $M$ the physical meaning of mass for a  spherical, \emph{holographic screen} with radius $r_\mathrm{h}$.  
The basic idea is that gravitational phenomena taking place in three-dimensional space can be projected on 
a two-dimensional ``~viewing screen~" with
no loss of information \cite{Susskind:1994vu}. The idea of holographic screen has been proposed in \cite{Verlinde} 
and it has mathematically been formulated in \cite{Chen:2010ay}: the holographic screen plays the role of ``~basic constituent~'' of space where the Newton potential is constant.''  Along this line of reasoning, the idea of holographic screen has been used also in the context of noncommutative inspired metric to derive compelling deviations to Newton's law \cite{Nicolini2010}. For what concerns the current discussion, however, we just need to recall that a special case of holographic screen is given by an event horizon where the entropy is maximized. 

Several remarks are in order:
\begin{itemize}
 \item It is easy to show that $M\ge \MP$ and equals the Planck mass only for $r_\mathrm{h}=\LP$.
 \item The line element (\ref{eq:metric-2}) admits a pair of horizons provided $M\ge \MP$.
The radii $r_\pm$ of the horizons are given by
\begin{equation}
 r_\pm = \LP^2\,\left(\, M \pm \sqrt{M^2 -\MP^2}\,\right)\label{hs}
\end{equation}
For $M=\MP$ the two horizons merge into a single (degenerate) null surface at $r_\pm =r_0=\LP$. 
For $M\gg\MP$ 
the outer horizon approaches the conventional value of the Schwarzschild geometry, \textit{i.e.}, $r_+\simeq 2M\LP^2$.
\item By inserting (\ref{mholo}) into (\ref{hs}) one finds
$r_+=r_\mathrm{h}$ , $r_-=\frac{\LP^2}{r_\mathrm{h}}$.
We see that the holographic screen surface coincides with the (outer) black hole horizon $r_+$, while
the inner Cauchy horizon has a radius which is always smaller or equal to the Planck length. 
This fact lets us circumvent the issue of potential blue shift instabilities \cite{ep1,ep2} (see for instance 
 recent analyses for noncommutative inspired \cite{cauchy,cauchy2} and other quantum gravity corrected metrics \cite{cauchy3,cauchy4})  because $r_-$ simply 
loses its physical meaning being not accessible to any sort of measurement process.
In what follows we can identify the holographic screen with the black hole outer horizon without distinguishing between the 
two surfaces any longer.
\item ``Light" objects, with  $M< \MP$, are ``particles'' rather than holographic screens. By particles
we mean localized lumps of energy of linear size given by the Compton wavelength $\lambda_C=1/M$, that can never collapse into 
a black hole. 
Rather they give rise to horizon-less metrics (see Fig. \ref{fig:hsc}) and cannot probe distances smaller than $\lambda_C$.  
The ``transition'' particle$\longrightarrow$ black
holes is discussed below in terms of critical \emph{surface density}.  
\end{itemize}
As a further analysis of this result, it is interesting to consider the \emph{surface energy density} of the holographic screen which is
defined as
\begin{equation}
 \sigma_{\mathrm{h}}\equiv \frac{M}{4\pi r_\mathrm{h}^2}=\frac{1}{8\pi\LP^2}\frac{r_+^2 +\LP^2}{r_+^3}\ .
\label{sigmah}
\end{equation}
From the above relation we see that $\sigma_\mathrm{h}$ is a monotonically decreasing function of the screen 
radius 
We notice that there exists a minimal screen encoding the physically maximum attainable energy density, 
\textit{i.e.} the Planck (surface) density:
\begin{equation}
 \sigma_h\left(\,r_+=\LP\,\right)= \frac{1}{4\pi\LP^3}=\frac{\MP}{4\pi\LP^2}.
\label{sigmapl}
\end{equation}
We stress that there is no physically meaningful ``~interior~'' for the minimal screen, \textit{i.e.} the ``volume'' 
of such an object is not even defined, in the sense that it can never be probed.  Thus, we can only consider energy per unit area, rather than per 
unit volume.
If we, formally, define a surface energy for a particle as
\begin{equation}
 \sigma_p\equiv \frac{M}{4\pi \lambda_C^2}=\frac{1}{4\pi \lambda_C^3}\label{sigmap}
\end{equation}
we see that the two curves (\ref{sigmah}) and (\ref{sigmap}) cross at $\lambda_C=\LP=r_+$. This result offers an additional interpretation for the Planck 
length which consistently turns to be the minimal size for a particle as well for a black hole (see Fig. \ref{fig:hsc}).  Accordingly,
the Planck density (\ref{sigmapl}) is the \emph{critical density} for a particle to collapse into
a black hole. This argument is usually formulated in terms of volume energy density having in mind
the picture of macroscopic body gravitationally collapsing under their own weight. From our holographic
vantage point, where ``surfaces'' are the basic dynamical objects, it is natural to reformulate 
this reasoning in terms of areal densities \cite{Susskind:1994vu}.   
In addition holography offers a way to circumvent potential conflicts between the mechanism of spontaneous 
dimensional reduction \cite{dr,dr2} and the UV self complete paradigm. If we perform the limit for $r\to 0$ 
the metric (\ref{eq:metric-2}) would apparently reduce into an effective two-dimensional spacetime
\begin{eqnarray}
ds^2\longrightarrow &-\left(\,1-2M\, r \,\right)dt^{2}
+\left( \, 1-2M\, r \, \right)^{-1}dr^{2}+\mathcal{ O}\left(r^2/\LP^2\right).
\end{eqnarray}
As explained in \cite{Mureika:2012fq}, this mechanism would lead the formation of lower dimensional black holes 
for length scales below the Planck length, in contrast with the predicted semi-classical regime of 
trans-Planckian black holes in four dimensions. However, contrary to the Schwarzschild metric that eventually 
reduces into dilaton gravity black holes when $r\simeq \LP$ (for reviews of the mechanism see \cite{dilaton1,dilaton2}), 
the presence of the holographic screen forbids the access to length scales $r<\LP$ and safely protects the 
arguments at the basis of the UV self complete quantum gravity.


\section{Thermodynamics, area quantization and mass spectrum.} 

\begin{figure}[t!]
\begin{center}
\includegraphics[width=0.6\textwidth]{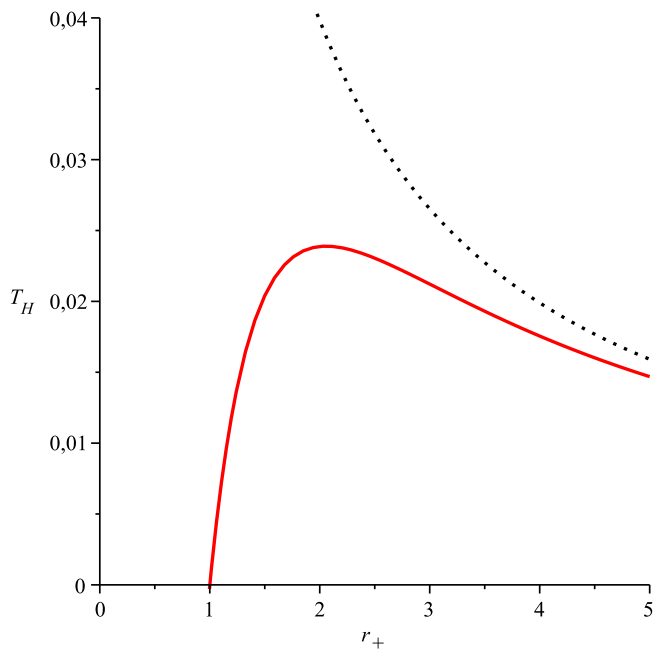}
\caption{\label{fig:thermo1}The solid curve represents the Hawking temperature $\T$ 
and as a function of the horizon radius $r_{+}$ 
in Planck units. The dotted curve represents the corresponding classical result in terms of the Schwarzschild metric.}
\end{center}
\end{figure}

\begin{figure}[t!]
\begin{center}
\includegraphics[width=0.6\textwidth]{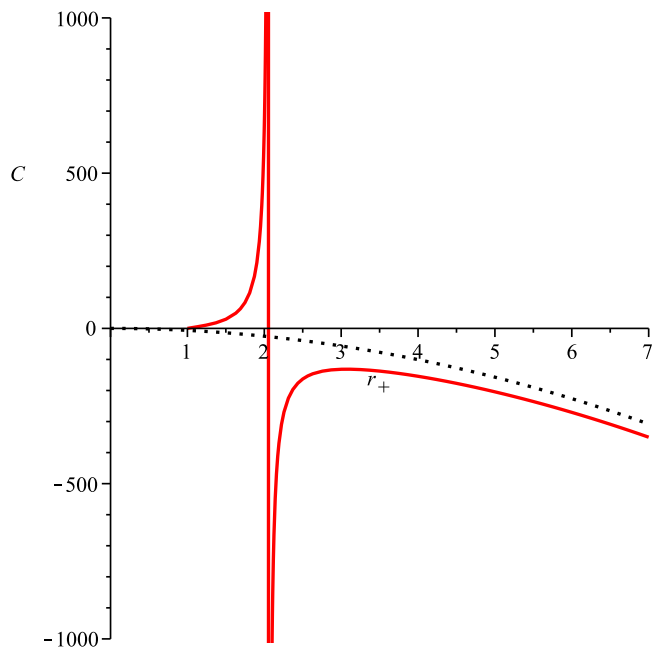}
\caption{\label{fig:thermo2}The solid curve represents  the black hole heat capacity $C$ 
as a function of the horizon radius $r_{+}$ 
in Planck units. The dotted curve represents the corresponding classical results in terms of the Schwarzschild metric.}
\end{center}
\end{figure}

In this section we would like to investigate the thermodynamics of the black hole described by (\ref{eq:metric-2}) 
and determine the relation between entropy and area of the event horizon.
It is customary to consider the area law for granted in any case, but this assumption
leads to an inconsistency with the Third Law of thermodynamics: extremal black holes
have zero temperature but non vanishing area. Here, we stick to the textbook definition
of thermodynamical entropy and not to more exotic quantity like R\'enyi, or entanglement
entropy.  To cure this flaw, we shall \emph{derive} the relation between entropy and
area from the First Law, rather than assuming it. 
The Hawking temperature associated to the metric (\ref{eq:metric-2}) can be calculated by evaluating the 
surface gravity  $\kappa$ 
\begin{equation}
\T=\frac{\kappa}{2\pi}=\frac{1}{4\pi}\left(\frac{dg_{00}}{dr}\right)_{r=r_+}  =\frac{1}{4\pi r_{+}}
\left(1-\frac{2\LP^2}{r^2_++\LP^2}
\right)\label{eq:temp}
\end{equation}
 while the heat capacity $C\equiv\partial U/\partial \T $ is 
\begin{equation}
C\equiv  \frac{\partial M}{\partial \T}
=-2\pi r_+\left(\frac{r_+^2-\LP^2}{\LP^2}\right)\frac{(r_+^2+\LP^2)^2}{r_+^4-4\LP^2 r_+^2-\LP^4}\ .\label{eq:heat}
\end{equation}
%
%
%
%
%
One can check that for large distances, \textit{i.e.}, $r_{+}\gg\LP$ both (\ref{eq:temp}) and (\ref{eq:heat}) coincide with 
the conventional results of the Schwarzschild 
metric, \textit{i.e.}, $\T\approx\frac{1}{4\pi r_{+}}$ and $C\approx-2\pi r_{+}^{2}$ (see Fig. \ref{fig:thermo1} and Fig. \ref{fig:thermo2}). 
On the other hand at Planckian scales, contrary to the standard result for which a Planckian black hole has a temperature 
$\T= \MP/8\pi $, we have 
that $\T\longrightarrow 0 $ as $r_+ \to r_0=\LP$ as expected for any extremal  configurations. 
This discrepancy with the classical picture is consistent with the genuine \emph{quantum gravitational} character  of the black hole and is reminiscent of the modified thermodynamics of noncommutative inspired black holes \cite{korea,india2}.  

The Hawking emission  is a semi-classical decay where gravity is considered just in terms of  a classical spacetime background. 
Such a semi-classical approximation 
conventionally breaks down as the Planck scale is  approached.
On the other hand for our metric, at $r_{+}=r_{\mathrm{M}}=\sqrt{2+\sqrt{5}}\LP\simeq 2.058\LP$  the temperature admits a 
maximum corresponding to a pole in the heat capacity. 
In the final stage of the evaporation, 
\textit{i.e.} $\LP<r_{+}<r_{\mathrm{M}}$, the heat capacity is positive, the Hawking emission  slows down and switches  
off at $r_{+}=\LP$. 
From a numerical estimate of the maximum temperature one finds 
$\T(r_{\mathrm{M}})=0.0239\MP$. This implies that the ratio temperature/mass is 
$\T/M<\T(r_{\mathrm{M}})/M_0\simeq 0.0239$. As a consequence, no relevant back 
reaction occurs during all the evaporation process and the metric can consistently describe the system 
``black hole + radiation" for all $r_+\geq\LP$.  

We can summarize the process with the following scheme:
 \begin{itemize}
  \item \emph{``~large~"}, far-from-extremality, black holes are  semi-classical objects which radiates thermally;
\item \emph{``~small~"}, quasi-extremal, black holes are quantum objects;
    \item $r=\rM$ is \emph{``~critical point~''} where the heat capacity diverges (see Fig. \ref{fig:thermo2}).
      Since $C>0$ for $r_0 < r_+ < r_M$ and $C< 0$ for $r_M < r_+ $, we conclude that a phase transition takes 
place from large thermodynamically unstable black holes to small stable black holes.  
   \end{itemize}

As a matter of fact, the black hole emission preceding the evaporation switching off 
(often called  ``SCRAM phase" \cite{Nicolini:2008aj}) might not be thermal. It has been argued that such a 
quantum regime might be characterized by discrete jumps towards the ground state \cite{gia2,Meade:2007sz}. 
To clarify the nature of this mechanism we proceed by studying the black hole entropy profile  and the related 
area quantization.
By integrating  the First Law, taking into account that no black hole can have a radius
smaller than $r_0=\LP$, \textit{i.e.}, 
\begin{equation}
S(r_{+})=\int_{r_0}^{r_+}\frac{dM}{\T}=\frac{\pi}{\LP^2}\left(\, r_+^2-\LP^2\,\right)+2\pi\ln\left(\, \frac{r_+}{\LP}\,\right)\ .
\label{eq:entropy_general}
\end{equation}
%
We can cast the entropy in terms of the area of the event horizon $\mathcal{A}_+\equiv 4\pi\, r_{+}^{2}$ as
\begin{equation}
S(\mathcal{A}_+)=\frac{\pi}{\mathcal{A}_0}\left(\, \mathcal{A}_+-\mathcal{A}_0\,\right)
+\pi\ln\left(\,\mathcal{A}_+/\mathcal{A}_0\, \right)
\label{eq:area-entropy}
\end{equation}
where $\mathcal{A}_0=4\pi\LP^2$ is the area of the extremal event horizon. We remark that the modifications to 
the Schwarzschild metric, encoded in our model, are in agreement with all the major approaches to quantum gravity, 
which universally foresee a logarithmic term as a correction to the classical area law. For brevity we recall that this is the case for string theory \cite{entropystring,entropystring2}, loop quantum gravity \cite{entropyloop1,entropyloop2,entropyloop3} and other results based on generic arguments \cite{entropy1,entropy2}, on the Cardy's formula \cite{entropy3}, conformal properties of spacetimes \cite{entropy4} and other mechanism for counting microstates \cite{entropy5,entropy6,entropy7}.
We can check that this is the case for the metric (\ref{eq:metric-2}) by performing the limit $r_+\gg\LP$ for (\ref{eq:area-entropy}) to obtain
\begin{equation}
S(\mathcal{A}_+)\approx\frac{\mathcal{A}_+}{4\LP^2}+\pi\ln\left(\frac{\mathcal{A}_+}{4\pi\LP^2}\right).
\end{equation}
Conversely for $r_+\to\LP$ the entropy vanishes, \textit{i.e.},
\begin{equation}
S(\mathcal{A}_+)\approx \frac{4\pi}{\LP}\left(\, r_+ - \LP\,\right) 
+ O\left(\, \left(\,  r_+ - \LP\,\right)^2\,\right).
\end{equation}
This result is consistent both with the Third Law of thermodynamics and the entropy statistical meaning.
The Planck size, zero temperature, black hole configuration is the unique ground state for  holographic screens. 
Thus, it is a  zero entropy state as there is only one way to  realize this configuration. 
To see this we promote the extremal configuration area to the  fundamental quantum of area.

\begin{equation}
\mathcal{A_{+}}\equiv \mathcal{A}_{n-1} =n\,\mathcal{A}_{0}=4\pi n\LP^2,
\end{equation}
where $\LP^2$ represents the basic information  pixel and $n=1,\, 2,\, 3 \ldots$ is the number of bytes.
\footnote{We borrow here the names of some units of digital information. In the present context, 
each byte consists of $4\pi$ bits. Each bit, represented by $\LP^2$ is the basic capacity of information 
of the holographic screen. In the analogy with the theory of information for which a byte represents 
the minimum amount of bits for encoding a single character of text, here the byte represents the minimum
 number of basic pixel $\LP^2$ for encoding the smallest holographic screen.} From the above condition one obtains
\begin{eqnarray}
&& r_{n-1}\equiv n^{1/2}\LP\ ,\\
&& M_{n-1}\equiv\frac{1}{2}\left(n^{1/2}+n^{-1/2}\right)\MP\ .
\end{eqnarray}
Consistently the ground state of the system is $r_{0}=\LP$ and $M_{0}=\MP$, 
while for $n\gg1$ one finds a continuous spectrum of values. This can be checked through the following relation
\begin{equation}
\Delta M_n\equiv M_{n}-M_{n-1}\sim\frac{1}{4}n^{-1/2}\MP.
\end{equation}
We notice that for $n\leq 4$ we are in the regime of positive heat
capacity $C>0$ and discrete mass spectrum, while for $n>4$ we
approach the semi-classical limit characterized by negative heat capacity
$C<0$ and continuous mass spectrum, \textit{i.e.}, $\Delta M_n/M_n\leq 1/12$. This confirms that 
at $r_{+}=r_{\mathrm{M}}$, the system undergoes a phase transition from a semi-classical regime to a 
 genuine quantum gravity regime. As a conclusion we have that large black holes decay thermally, while 
small objects decay quantum mechanically, by emitting quanta of energy (for a recent phenomenological analysis 
of such kind of decay see \cite{calmet}).
The end-point of the decay is a Planck mass,  holographic screen.

The quantization of the area of the holographic screen lets us disclose further features of the 
informational content of the holographic screen. We have that the surface density can be written as
\begin{equation}
\sigma_\mathrm{h}(n)=\frac{1}{2}\left(\frac{1}{n^{1/2}}+\frac{1}{n^{3/2}}\right)\frac{\MP}{4\pi\LP^2}
\end{equation}
while the entropy reads $ S(n)=\pi\left(\, n+\ln(n)-1\, \right)$.
From this relations we learn that while the entropy increases with the number $n$ of bytes, 
the surface density decreases. This confirms that the extremal configuration is nothing but a 
single byte, zero-entropy, Planckian density holographic screen.

\section{Discussion and conclusions}
 
In this paper we have presented a neutral non-spinning black hole geometry admitting an extremal configuration whose mass and radius 
coincide with the Planck units. 
We have reached this goal by suitably modelling a stress tensor able to accommodate both the particle and black hole configurations, 
undergoing a transition at the Planck scale. 
We showed that the horizon of the degenerate black hole 
represents the minimal holographic screen, within which we cannot access to any information about the
 matter-energy content of spacetime. 
 

We showed that a generic holographic screen is described  in terms of the outer horizon of 
the metric  (\ref{eq:metric-2}), while the inner horizon lies within the prohibited region, \textit{i.e.}, 
inside the minimal holographic screen. The whole scheme fits into the 
gravity self-completeness scenario. 
For sub-Planckian energy scales one has just a quantum particle able to probe at the most distances of the order of its 
Compton wavelength. By increasing the degree of 
compression of the particle, one traverses the  Planck scale where a collapse into a black hole occurs, 
before probing a semi-classical regime at trans-Planckian energies. 
The virtual curvature singularity of the geometry in $r=0$ is therefore wiped out since in such a 
context sub-Planckian lengths have no physical meaning. From this vantage point spacetime stops 
to exists beyond the Planck scale as there is no physical way to access this regime. 
Thus, the curvature singularity problem is ultimately resolved by giving up the very concept
of spacetime at sub-Planckian length scales.     

The study of the associated thermodynamic quantities confirmed that at trans-Planckian energies black holes radiate thermally 
before undergoing a phase transition to smaller, 
quantum black holes. The latter decay by emitting a discrete spectrum of quanta of energy and reach the ground state of the 
evaporation corresponding to the minimal holographic 
screen.  We came to this conclusion by quantizing the black hole horizon area in terms of the minimal holographic screen 
which actually plays the role of a basic information byte. 
We showed that in the thermodynamic limit, the area law for the black hole entropy acquires a logarithmic correction in
agreement with all the major quantum gravity formulations. 

In conclusion, we stress that the line element (\ref{eq:metric-2}) not only captures the basic features of  more 
``~sophisticated~'' models of quantum gravity improved 
black holes (\textit{e.g.} noncommutative geometry inspired black holes \cite{Nicolini:2008aj}, loop quantum gravity black holes \cite{lbh3,lbh4}, 
asymptotically safe gravity black holes \cite{br,br2} and other studies about collapses in quantum gravity \cite{lbh1,lbh2}), but overcomes some of their current weak points: 
specifically there is no longer any concern for potential Cauchy instabilities or for conflicts between 
the gravity self-completeness and the Planck scale spontaneous dimensional reduction mechanism, as well 
as, the scenario of the terminal phase of the evaporation for static, non-rotating, neutral black holes.  
In addition, for its compact form the new metric allows straightforward analytic calculations and opens 
the route to testable predictions.

\end{document}